\documentclass[twocolumn,showpacs,preprintnumbers]{revtex4}
\usepackage{dcolumn}
\usepackage{bm}
\usepackage{amsmath}
\usepackage{graphicx}
\usepackage{mathrsfs}
\usepackage{amssymb}
\usepackage{amsfonts}

\begin{document}

\title{Generation of two-mode entanglement via atomic coherence
created by non-resonant dressed-state transitions}
\author{Jinhua Zou,$^{1,2}$ \footnote{E-mail:jhzou@yangtzeu.edu.cn} Dahai Xu $^{2}$ and Huafeng Zhang,$^{2}$}
\affiliation{$^{1}$State Key Laboratory of Magnetic Resonance and Atomic and
Molecular Physics, Wuhan Institute of Physics and Mathematics, Chinese
Academy of Sciences, Wuhan, 430071,China\\
\newline
$^{2}$ The School of Physical Science and Technology, Yangtze University, Jingzhou, 434023, China}

\begin{abstract}
Two-mode squeezing and entanglement is obtained in a atom-cavity system
cosisting a three-level atom and a two-mode cavity with driving laser
fields. Here non-resonatn dressed-state transitions between the cavity modes
and atom are used to get the effective Hamiltonian.
\end{abstract}
\maketitle

\draft

\vskip 2cm

\narrowtext

\newpage

\section{Introduction}

Quantum entanglement is regarded as the magic part of mechanics and has
became one central part in quantum information and quantum computation.
Recently, continuous-variable (CV) entanglement has attracted much great
interest in the application of entanglement-based quantum information
processing \cite{1}. Thus generation continuous-variable entanglement have gained
much attention since they can provide useful entanglement sources for
quantum teleportation [2], quantum dense coding [3] and quantum swapping
[4]. There are many interaction system candidates to generate CV
entanglement such as cavity QED system [5-9], nondegenerate parametric
down-conversion optical crystals [10,11], NV centers system [12,13], and so
on.

As for cavity QED system, it has been a good candidate to investigate the
quantum interaction in physics and quantum mechanics. Many schemes have been
presented to generate entangled states. For example, the scheme for
generating CV entangled light via correlated emission laser has been
proposed [14]. They obtain a two-mode entangled light with large photon
number. The scheme to generate CV entanglement via a single atom laser has
also been presented [15,16]. Among the entanglement generating schemes, some
schemes use atomic coherence to build the entanglement between various
cavity modes [17,18]. The atomic coherence can be input or created by
driving related transitions. In these schemes, atomic coherence induces
entanglement somewhat. One scheme is proposed in Ref. [19],  a two-step
operation to get highly entangled two-mode light by using the atoms as a
reservior is proposed. They use the dressed states of the driving field and
choose resonant transition for the two cavity modes and the
squeeze-transformed modes successively interact with the dressed atoms. One
of the limitations in this scheme are as following. First, the two-step
procedure and different initial state preparation make the experimental
setup complicated. It is useful to investigate the scheme that has one-step
procedure and simple initial state preparation.

Motivated by this, we propose a scheme to extend the atomic system to a
three-level atom, which has an auxiliary driving transition to generate the
two-mode entanglement light. By introducing the auxiliary driving transition
and choosing non-resonant atom-cavity interaction, the dressed atomic-cavity
interaction can be an effective two-mode squeezing interaction, which can be
used to generate two-mode entangled light.

\section{Model and equations}

The system we consider is a three-level $\Lambda $ atomic system locates in
a two-mode optical cavity with drivings. Two classical laser fields with
Rabi frequency $\Omega _{j}$ and frequency $\omega _{d,j\text{ }}$resonantly
drive the two atomic transitions $|j\rangle \leftrightarrow |3\rangle $,
respectively for $j=1$, $2$. The two cavity modes with coupling constant $%
g_{n}$ and frequency $\omega _{c,n}$ coupling transitions $|2\rangle
\leftrightarrow |3\rangle $ ( $n=1,2$). The whole system can be described by
the following Hamiltonian ($\hbar =1$)
\begin{eqnarray}
H &=&H_{\Omega }+H_{c}  \nonumber \\
H_{\Omega } &=&-\sum_{j=1,2}\Omega _{j}(|3\rangle \langle j|+|j\rangle
\langle 3|)  \nonumber \\
H_{c} &=&\sum_{n=1,2}\delta _{n}a_{n}^{\dagger
}a_{n}+\sum_{n=1,2}[(g_{n}a_{n}\sigma _{32}+g_{n}^{*}a_{n}^{\dagger }\sigma
_{23})  \label{1}
\end{eqnarray}
where $\Delta _{j}=\omega _{3j}-\omega _{d,j\text{ }}$( $j=1,2$) and $\delta
_{n}=\omega _{c,n}-\omega _{d,2\text{ }}$with $\omega _{3j}$ are atomic
resonant frequency for transition $|j\rangle \leftrightarrow |3\rangle $ ( $%
j=1,2$). Atomic operator $\sigma _{jk}$ are flip operators when $j\neq k$
and are population operators when $j=k$. $a_{n}$ ($a_{n}^{\dagger }$) is the
creation (annihilation) operator for the $n$th cavity mode.

Assume that the laser fields are much strong, we first calculate the dressed
states created by part Hamiltonian $H_{\Omega }$. The eigenvalues are $%
\lambda _{0}=0$, $\lambda _{\pm }=\pm \sqrt{\Omega _{2}^{2}+\Omega _{1}^{2}}%
=\pm \Omega _{e}$, and related eigenstates are

\begin{eqnarray}
|0\rangle &=&\cos \theta |2\rangle -\sin \theta |1\rangle  \nonumber \\
|+\rangle &=&\frac{1}{\sqrt{2}}|3\rangle +\frac{\sin \theta }{\sqrt{2}}%
|2\rangle +\frac{\sin \theta }{\sqrt{2}}|1\rangle  \label{2} \\
|-\rangle &=&-\frac{1}{\sqrt{2}}|3\rangle +\frac{\sin \theta }{\sqrt{2}}%
|2\rangle +\frac{\sin \theta }{\sqrt{2}}|1\rangle  \nonumber
\end{eqnarray}
in which $\sin \theta =\frac{\Omega _{1}}{\Omega _{e}}$. Thus in the
eigenstates basis, the ralations hold $H_{\Omega }=$ $\Omega _{e}(\sigma
_{++}-\sigma _{--})$ and $\sigma _{32}=\frac{1}{\sqrt{2}}\cos \theta (\sigma
_{+0}-\sigma _{-0})+\frac{\sin \theta }{2}(\sigma _{++}-\sigma _{-+}+\sigma
_{+-}-\sigma _{--})$. Then make a unitary transformation $%
e^{iH_{0}t}H_{c}e^{-iH_{0}t}$ with $H_{0}=H_{\Omega }+\sum_{n=1,2}\delta
_{n}a_{n}^{\dagger }a_{n}$ and assume that the detunings $\delta
_{1}=2\Omega _{e}-d$ and $\delta _{2}=2\Omega _{e}+d$, where $d>0$ is a
small quantity compared with $2\Omega _{e}$ but is still large when compared
with the effective coupling strength (which will show later). Because of the
strong driving condition, the unitary transformation will contain fast
osicilating terms $e^{\pm i\Omega _{e}t}$ and $e^{\pm 2i\Omega _{e}t}$, by
neglecting theses terms, i.e., when secular appoximation is made, the cavity
related Hamiltonian now has the form

\begin{equation}
H_{c1}=(G_{1}a_{1}-G_{2}^{*}a_{2}^{\dagger })\sigma
_{+-}e^{idt}+(G_{1}^{*}a_{1}-G_{2}a_{2})\sigma _{-+}e^{-idt}  \label{3}
\end{equation}
in which $G_{1}=\frac{g_{1}}{2}\sin \theta $, and $G_{2}=\frac{g_{2}}{2}\sin
\theta $. If $d$ $\gg |G_{1}|,|G_{2}|$ is fulfilled, an effective
Hamiltonian $H_{e}$ can be obtained as
\begin{eqnarray}
H_{e} &=&\lambda (a_{1}a_{2}+a_{1}^{\dagger }a_{2}^{\dagger })(\sigma
_{++}-\sigma _{--})  \nonumber \\
&&+(\frac{|G_{1}|^{2}}{d}a_{1}^{\dagger }a_{1}+\frac{|G_{2}|^{2}}{d}%
a_{2}^{\dagger }a_{2})(\sigma _{++}-\sigma _{--})  \label{4} \\
&&+\lambda _{++}\sigma _{++}-\lambda _{--}\sigma _{--}  \nonumber
\end{eqnarray}
where $\lambda =\frac{G_{1}G2}{d}$ is the effective coupling between the two
cavity modes, and $\lambda _{++}=$ $\frac{|G_{1}|^{2}}{d}$, $\lambda _{--}=%
\frac{|G_{2}|^{2}}{d}$. Assume that the two cavity modes are initially in
their vacuum states $|0_{1}\rangle |0_{2}\rangle $, then the terms related
to $a_{j}^{\dagger }a_{j}$ ($j=1,2$) have no contribution to the evolution
and can be omitted. Thus the final effective Hamiltonian is

\begin{eqnarray}
H_{eff} &=&\lambda (a_{1}a_{2}+a_{1}^{\dagger }a_{2}^{\dagger })(\sigma
_{++}-\sigma _{--})  \nonumber \\
&&+\lambda _{++}\sigma _{++}-\lambda _{--}\sigma _{--}  \label{5}
\end{eqnarray}

Next we will show how to obtain squeezing and entanglement from this
effective Hamiltonian for the two cavity modes. If the initial state of the
whole system is $|\psi \left( 0\right) \rangle =|+\rangle |0_{1}\rangle
|0_{2}\rangle $, then the state will evolve as $|\psi \left( t\right)
\rangle =e^{-iH_{e}t}|\psi \left( 0\right) \rangle $ to the following state
\begin{equation}
|\psi \left( t\right) \rangle =e^{-i\lambda _{++}t}|+\rangle e^{-i\lambda
(a_{1}a_{2}+a_{1}^{\dagger }a_{2}^{\dagger })t}|0_{1}\rangle |0_{2}\rangle
\label{6}
\end{equation}
Thus if the whole atom is intially in state $|+\rangle |0_{1}\rangle
|0_{2}\rangle $, then after time $t$, the atom will still in its state $%
|+\rangle $, but the two cavity modes will evolve into the following state

\begin{equation}
|\varphi \left( t\right) \rangle _{c}=e^{-i\lambda
t(a_{1}a_{2}+a_{1}^{\dagger }a_{2}^{\dagger })}|0_{1}\rangle |0_{2}\rangle
\label{7}
\end{equation}

It is obvious that $|\varphi \left( t\right) \rangle _{c}$ is a two-mode
squeezed vacuum state. Let us define the quadrature operators of the two
cavity modes which will be used to judge squeezing and entanglement as $%
x_{j}=\frac{a_{j}+a_{j}^{\dagger }}{\sqrt{2}}$, $p_{j}=-i\frac{%
a_{j}-a_{j}^{\dagger }}{\sqrt{2}}$, and $u=x_{1}+x_{2}$, $\upsilon
=p_{1}-p_{2}$. The quanties $\langle (\Delta u)^{2}\rangle =\langle
u^{2}\rangle -\langle u\rangle ^{2}$ and $\langle (\Delta \upsilon
)^{2}\rangle =\langle \upsilon ^{2}\rangle -\langle \upsilon \rangle ^{2}$
are needed to be calculated. According to the entantanglement criterion
[20], if $M=\langle (\Delta u)^{2}+(\Delta \upsilon )^{2}\rangle <2$, then
the two cavity modes are entangled. Using the state $|\varphi \left(
t\right) \rangle _{c}$ in Eq. (7), the following relations hold $\langle
u\rangle =\langle \upsilon \rangle =0$, $\langle a_{1}^{\dagger
}a_{1}\rangle =\langle a_{2}^{\dagger }a_{2}\rangle =\sinh (\lambda t)$ and $%
\langle a_{1}a_{2}\rangle =-i\cosh (\lambda t)\sinh (\lambda t)$. Thus $%
M=2+(e^{\lambda t}-e^{-\lambda t})^{2}=2\left( e^{2\lambda t}+e^{-2\lambda
t}\right) $. If $\lambda =i|\lambda |$ ($|\lambda |\neq 0$), then $M=2\cos
(2\lambda t)<2$, that is to say if we choose $g_{1}=|g|$, $g_{2}=i|g|$, then
$\lambda =\frac{G_{1}G2}{d}=i\frac{|g|^{2}}{d}$, and entanglement criterion $%
M<2$ always holds. Thus two cavity modes are entangled states. Actually the
state $|\varphi \left( t\right) \rangle _{c}$ is a two-mode entangled state
and is an ideally entangled state when $\lambda =i|\lambda |$ ($|\lambda
|\neq 0$) holds.

\section{Conclusion}

In conclusion, we have presented a scheme to generate two-mode entangled
state in a cavity QED system by using the atomic coherence created by the
two strong classical driving fields. The result shows that when we
appropriately adjusting the coupling strengths of the two cavity modes, a
two-mode squeezed vacuum state which is a two-mode entangled state is
generated.\\
\textbf{Acknowledgments}

This work is supported by the Scientific Research Plan of the 281 Provincial
Education Department in Hubei (Grant No. Q20101304 ) and NSFC under Grant
No. 11147153.

\end{document}